\documentclass[10pt,conference]{IEEEtran}
\usepackage{epsfig,rotating,setspace,latexsym,amsmath,epsf,amssymb,amsfonts,bm,theorem,cite,caption,subcaption,enumerate,longtable,accents,float}
\usepackage{algorithm,algorithmic,graphicx,epsf,authblk,epstopdf,url,xcolor, soul,multirow,bbm}
\usepackage{mathtools,comment}
\usepackage[center]{qtree}
\usepackage{tree-dvips}
\usepackage[linguistics]{forest }

\newtheorem{theorem}{Theorem}

\newtheorem{remark}{Remark}
\newtheorem{lemma}{Lemma}

\newenvironment{Proof}[1]{\medskip\par\noindent{\bf Proof:\,}\,#1}{{\mbox{\,$\blacksquare$}\par}}

\IEEEoverridecommandlockouts

\newcommand{\cw}{{\mathcal{W}}}
\newcommand{\cs}{{\mathcal{S}}}
\newcommand{\ce}{{\mathcal{E}}}

\DeclareMathOperator{\diag}{diag}

\allowdisplaybreaks

\title{Private Membership Aggregation}
\author{Mohamed Nomeir \qquad Sajani Vithana \qquad Sennur Ulukus\\
	\normalsize Department of Electrical and Computer Engineering\\
	\normalsize University of Maryland, College Park, MD 20742 \\
	\normalsize \emph{mnomeir@umd.edu} \qquad \emph{spallego@umd.edu} \qquad \emph{ulukus@umd.edu}}
 
\begin{document}

\maketitle

\begin{abstract}
We consider the problem of private membership aggregation (PMA), in which a user counts the number of times a certain element is stored in a system of independent parties that store arbitrary sets of elements from a universal alphabet. The parties are not allowed to learn which element is being counted by the user. Further, neither the user nor the other parties are allowed to learn the stored elements of each party involved in the process. PMA is a generalization of the recently introduced problem of $K$ private set intersection ($K$-PSI). The $K$-PSI problem considers a set of $M$ parties storing arbitrary sets of elements, and a user who wants to determine if a certain element is repeated at least at $K$ parties out of the $M$ parties without learning which party has the required element and which party does not. To solve the general problem of PMA, we dissect it into four categories based on the privacy requirement and the collusions among databases/parties. We map these problems into equivalent private information retrieval (PIR) problems. We propose achievable schemes for each of the four variants of the problem based on the concept of cross-subspace alignment (CSA). The proposed schemes achieve \emph{linear} communication complexity as opposed to the state-of-the-art $K$-PSI scheme that requires \emph{exponential} complexity even though our PMA problems contain more security and privacy constraints.
\end{abstract}

\section{Introduction}
Multi-party computation (MPC) is used in a wide range of applications such as secure voting, privacy-preserving data analysis, collaborative machine learning, secure social networks, etc \cite{recent}. Private set intersection (PSI) is one of the most fundamental multi-party computations \cite{psi_oblivious_transfer_2, psi_permutation_hashing, psi_adversaries, practical-multiparty_psi_symmetric_key,  psi_oblivious_transfer, wang_psi, Wang_multiparty_PSI}. In PSI, there are multiple parties, each storing a set of elements coming from an alphabet. It is required to find the intersection of the sets of all parties without leaking any information about the remaining elements in each party beyond the intersection. \cite{wang_psi} formulates the two-party PSI problem from an information-theoretic point of view, finds the optimal download cost and proposes an optimum achievable scheme. \cite{Wang_multiparty_PSI} considers the multi-party version of PSI, determines the optimal download cost and gives a capacity-achieving scheme. The schemes in \cite{wang_psi, Wang_multiparty_PSI} are based on concepts from information-theoretic private information retrieval (PIR). 

A new variation of the multi-party PSI problem, called $K$-PSI, is recently introduced in \cite{yehia_federated_K_PSI}. In $K$-PSI, there are $M$ parties storing arbitrary sets of elements out of an alphabet. A user wishes to know if a certain element is repeated $K$ times or not among the $M$ parties. In this problem, the parties do not want to leak any information about their datasets to the other parties or to the user; the user should not learn which parties contain the queried element and which parties do not; and the parties should not learn any information about the element being queried. In \cite{yehia_federated_K_PSI}, a scheme is designed to solve the $K$-PSI problem with an exponential communication complexity, i.e., $\mathcal{O}(M^K(K-1))$. The same complexity can be achieved with weaker privacy using the existing schemes on PIR-based-PSI \cite{wang_psi, Wang_multiparty_PSI} if we allow each party to have $N$ databases. This motivates to look at $K$-PSI and its extensions through the lens of PIR, as it provides the elemental privacy and security requirements of any multi-server system \cite{c_pir, c_spir, pir_survey, salim_singleserver_sideinfo,banawan_pir_mdscoded, byzantine_tpir, csa, sajani_fed_learning_sparse, sajani_federated_learning, sajani_pruw,sajani_rateprivacystorage}. 

\begin{figure}
\centering
\includegraphics[width=0.8\columnwidth]{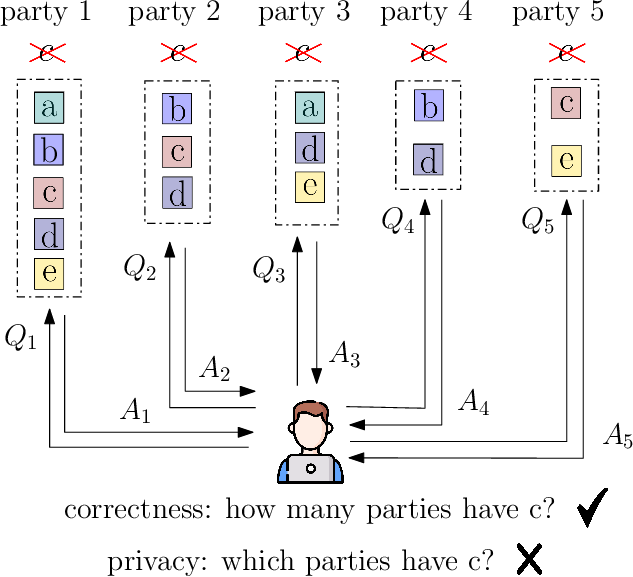}
\caption{Private membership aggregation (PMA) system model.}
\label{sys_1}
\vspace*{-0.4cm}
\end{figure}

In this paper, we generalize the problem of $K$-PSI by computing the \emph{exact number} of parties storing a certain element, without revealing the user any information about the elements stored in each party, and without letting the databases know which element is being checked. In addition, we do not allow the user to know which parties have the required element and which do not. We coin this problem as private membership aggregation (PMA); see Fig.~\ref{sys_1}. This is a \emph{fine-grained} version of $K$-PSI, as instead of asking if an element is repeated more than $K$ times in the parties, we ask how many times an element is repeated in the parties. The main applications of PMA include multiple identity detection and anomaly detection. For example, in the health insurance industry, companies want to make sure that a person with a certain social security number does not have another account in another company. Another application is to check the validity of certain information by making sure that it exists in some of the other parties as well. 

\begin{figure}
\centering
\includegraphics[width=\columnwidth]{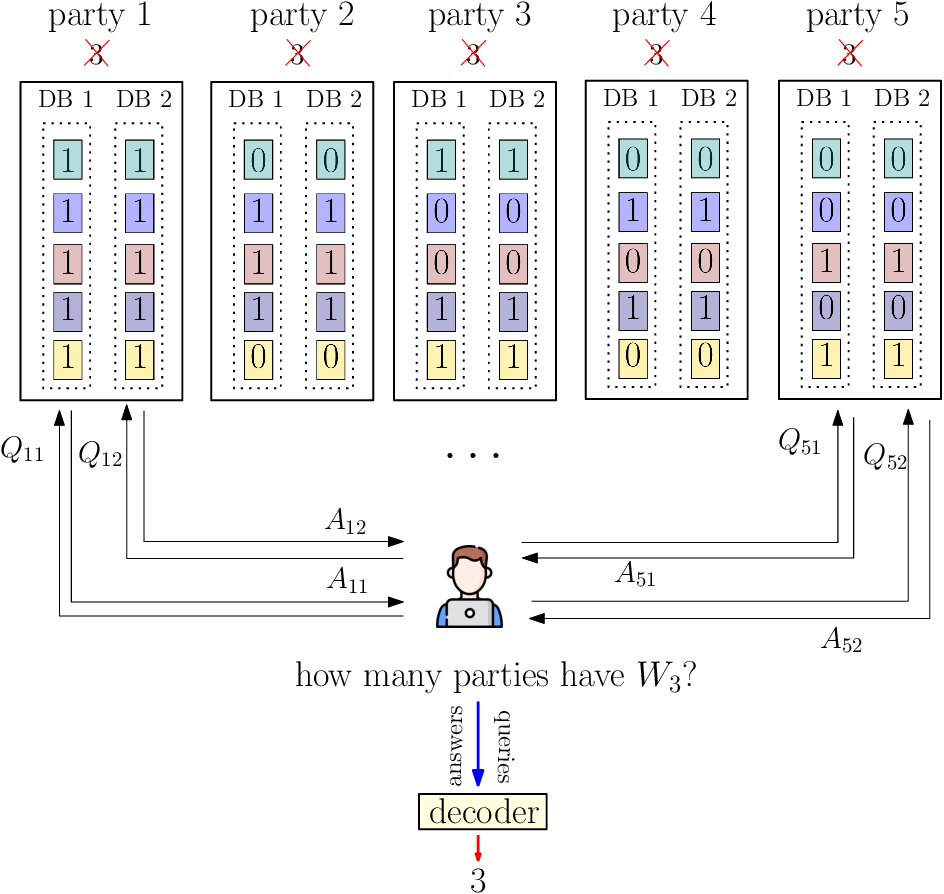}
\caption{PMA model: from elements to incidence vectors.}
\label{sys_2}
\vspace*{-0.4cm}
\end{figure}

In this work, we consider four different variants of PMA, described by the cases where: 1) different parties are allowed to eavesdrop on the answers from other parties, 2) the user is not allowed to learn any information about the elements other than what is being checked, which is coined as symmetric PMA (SPMA), 3) certain subsets of databases within each party are allowed to collude (type I collusion), and 4) certain subsets of parties are allowed to collude (type II collusion). We formulate each problem in the context of PIR, and use concepts from cross subspace alignment (CSA) \cite{csa} to solve each problem. We modify the basic CSA scheme to achieve the additional privacy requirement in SPMA by using a masking technique to keep the user from obtaining any information on the elements other than what is checked. We provide schemes that perform each variant of PMA stated above with a \emph{linear} communication complexity, which is a significant improvement compared to the complexity required by the existing state-of-the-art $K$-PSI schemes, which is \emph{exponential}. 

\section{Problem Formulation}\label{problem_formulation}
We consider $M$ parties, each containing $N$ servers. There are $E$ elements in total in the universal set each of which can be mapped to a separate message $W_k$, $k\in\{1,\dotsc,E\}$; see incidence vectors in Fig.~\ref{sys_2}. Each message $W_k$ has a probability $p_{k}$ to be in the message set of any given party, i.e.,
\begin{align}
    \mathbb{P}(W_k \in \mathcal{P}_i) = p_{k}, \quad i \in [M], ~k \in [E],
\end{align}
where $\mathcal{P}_i$ is the set of messages in the $i$th party. Each message $W_i$ is generated uniformly at random, independent of other messages, and independent from the shared common randomness between the parties, i.e., 
\begin{align}
H(W_{[E]},\cs) = E H(W_1)+ H(\cs),
\end{align}
where $\cs$ is the shared randomness among the parties. 

Each party wishes to keep their message contents $\mathcal{P}_i$ and the indices of the messages available at their datasets hidden from other parties, i.e., for each $i,j \in [M]$, $i\neq j$,
\begin{align}\label{parties_sec_1}
    I(\mathcal{P}_i; \ce_{ij}|\mathcal{P}_j,\ce_{1j},\ldots,\ce_{i-1,j},\ce_{i+1,j},\ldots,\ce_{Mj})&=0,\\
I(\mathcal{U}_i;\ce_{ij}|\mathcal{P}_j,\ce_{1j},\ldots,\ce_{i-1,j},\ce_{i+1,j},\ldots,\ce_{M,j})&=0,\label{parties_sec_2} 
\end{align}
where $\mathcal{U}_i = \bigcup_{j=1}^E\{\mathbbm{1}(W_j \in \mathcal{P}_i)\}$ and $\ce_{ij}$ is all possible communications between the $i$th and the $j$th parties.

The user chooses an index $\theta \in [E]$ uniformly at random and wishes to compute how many parties store $W_{\theta}$ by sending a query $Q^{\theta}_{ij}$ to the $j$th database in the $i$th party, which satisfies,
\begin{align}\label{PSI_1}
    I(\theta; Q_{ij}^{\theta}| \mathcal{P}_i) = 0, \quad i \in [M], ~j \in [N].
\end{align}

After receiving the queries, each database replies truthfully with an answer string $A_{ij}^{\theta}$ which is a deterministic function of the received query $Q_{ij}^{\theta}$, the messages available at each party $\mathcal{P}_i$, and shared randomness between parties $\cs$, i.e., 
\begin{align}
    H(A_{ij}^{\theta}| Q_{ij}^{\theta}, \mathcal{P}_i,\cs ) = 0, \quad i \in [M], ~j \in [N].
\end{align}

If the $i$th party is eavesdropping on $Y_{i\ell}$ links in the $\ell$th party, then, party $i$ should not be able to obtain any information on the message index being checked, contents of the $\ell$th party, or the required answer by the user, $\kappa^{\theta} = \sum_{i \in [M]}\mathbbm{1}(W_{\theta} \in \mathcal{P}_i)$, i.e., for $i,\ell \in [M]$, $i\neq \ell$, $\kappa^{\theta} \in \{0,\ldots,M\}$, $\theta \in [E]$,
\begin{align}
I(\theta; A^{\theta}_{\ell \mathcal{Y}_{\ell}},Q^{\theta}_{\ell \mathcal{Y}_{\ell}}|\cs, \mathcal{P}_i)&= 0,\\
I(\mathcal{U}_{\ell}; A^{\theta}_{\ell \mathcal{Y}_{\ell}},Q^{\theta}_{\ell \mathcal{Y}_{\ell}}|\cs, \mathcal{P}_i)&= 0,\\
 I(\kappa^{\theta}; A^{\theta}_{\ell \mathcal{Y}_{\ell}},Q^{\theta}_{\ell \mathcal{Y}_{\ell}}|\cs, \mathcal{P}_i)&= 0, 
\end{align}
where $\mathcal{Y}_{\ell}$ is the set of databases in party $\ell$ whose communication links are eavesdropped on by party $i$, such that $|\mathcal{Y}_{\ell}| \leq \max_i (Y_{i\ell})$. Given the answer strings from all parties, the user can apply a decoding scheme that generates the required answer with no error, i.e.,  
\begin{align}
    \hat{\kappa}^{\theta} = g(A_{ij}^{\theta},Q_{ij}^{\theta},~ i \in [M], ~j \in [N]), \quad \theta \in [E],
\end{align}
where $g$ is the decoding scheme, and
\begin{align}
    \mathbb{P}(\hat{\kappa}^{\theta} = \kappa^{\theta}) = 1.
\end{align}

In addition, PMA requires that the user should not be able to infer any information about the parties containing the required message, except by random guessing, i.e., no information about the locations of the required message $W_{\theta}$ can be extracted from the answer strings, i.e., for $ \theta \in [E], ~\kappa \in \{0, \dots, M\}, ~i \in [M],$
\begin{align}\label{PSI_2}
    \mathbb{P}(W_{\theta} \in \mathcal{P}_i| A^{\theta}, \kappa^{\theta}=\kappa) = \frac{\kappa}{M},
\end{align}
where $A^{\theta} = \bigcup_{i,j} \{A_{ij}^{\theta}\}$. Moreover, for any $\{i_1,\dotsc,i_n\}\subset[M]$ such that $n\leq\kappa$, 
\begin{align}\label{PSI_3}
    \mathbb{P}(W_{\theta} \in \{\mathcal{P}_{i_1},\ldots, \mathcal{P}_{i_n}\}| A^{\theta}, \kappa^{\theta}=\kappa) = \frac{{\kappa \choose n}}{{M \choose n}}.
\end{align}

Finally, in SPMA, it is required that no information about the messages other than the one being checked is allowed to leak to the user, i.e., for $\theta \in [E]$,
\begin{align}\label{fed_PSI_1}
I(\mathcal{W}_{\theta^C};A^{\theta}|Q^{\theta},\kappa^{\theta}) &= 0,\\
I(\Gamma^{\theta};A^{\theta}|Q^{\theta},\kappa^{\theta}) &= 0,\label{fed_PSI_2}
\end{align}
where $\cw_{\theta^C} = \{W_i: i \in [E], i \neq \theta\}$, $Q^{\theta} = \bigcup_{i,j} \{Q_{ij}^{\theta}\}$, and $\Gamma^{\theta} = \bigcup_{i=1}^M \bigcup_{j=1\atop{j\neq \theta}} ^E \{\mathbbm{1}(W_j \in \mathcal{P}_i)\}$. 

A scheme that satisfies \eqref{parties_sec_1}-\eqref{PSI_3} is called a PMA scheme, and a PMA scheme that satisfies \eqref{fed_PSI_1}, \eqref{fed_PSI_2} is called an SPMA scheme. The download cost $D$ of any of these schemes is  
\begin{align}
    D = \mathbb{E}\bigg[\sum_i \sum_j H(A_{ij}^{\theta})\bigg],
\end{align}
where the expectation is taken over $\theta$.

We further separate the problem into two types, namely, 1) collusions within the databases in each party, i.e., the parties do not collude but the databases within each party are allowed to collude (type I collusion), and 2) collusions among the parties, i.e., the databases within each party are colluding with other databases from other parties (type II collusion). 

\section{Main Results}\label{main_results}
\begin{theorem}\label{optimal downlaod cost k-psi-1}
     Consider a PMA system with type I collusions consisting of $M$ parties, each of which has $N$ databases with any $T$ of them colluding. Each party is allowed to eavesdrop on $Y$ links of the other parties. The optimal download cost of this case $D^*_{\text{PMA-I}}$ must satisfy,
     \begin{align}
         D^*_{\text{PMA-I}} \leq M (\max(T,Y)+1),
     \end{align}
     with $N \geq \max(T,Y)+1$.
\end{theorem}

\begin{theorem}\label{optimal download cost fed-k-psi-1}
    For the same setting as in Theorem~\ref{optimal downlaod cost k-psi-1} with the additional condition of symmetric privacy, the optimal download cost of SPMA $D^*_{\text{SPMA-I}}$ must satisfy,
    \begin{align}
        D^*_{\text{SPMA-I}} \leq M (\max(T,Y)+1),
    \end{align}
    with $N \geq \max(T,Y)+1$.
\end{theorem}

\begin{theorem}\label{optimal download cost fed-k-psi-2}
     Consider a PMA system with type II collusions consisting of $M$ parties, each of which has $N$ databases. All databases in any $T$ out of the $M$ parties can collude, and the $i$th party is able to listen to $Y_i$ links of any other party. The optimal download cost of non-symmetric and symmetric variants of this case $D^*_{\text{PMA-II}}$ and $D^*_{\text{SPMA-II}}$ must satisfy,
     \begin{align}
         D^*_{\text{PMA-II}} &\leq N+\max(TN,Y_1,\ldots,Y_M) + 1,\\
         D^*_{\text{SPMA-II}} &\leq N+\max(TN,Y_1,\ldots,Y_M) + 1,
     \end{align}
     with $MN \geq N+\max(TN,Y_1,\ldots,Y_M) + 1$.
\end{theorem}

\begin{remark}
    The bounds on the download costs in Theorems~\ref{optimal downlaod cost k-psi-1}-\ref{optimal download cost fed-k-psi-2} do not depend on the number of messages in the system $E$.  
\end{remark}

\begin{remark}
    The achievable schemes for Theorems~\ref{optimal downlaod cost k-psi-1} and~\ref{optimal download cost fed-k-psi-1} result in the same download cost for PMA type I and SPMA type I. This is because the modified CSA scheme that achieves symmetric privacy does not require any additional downloads. 
\end{remark}

\begin{remark}
    The related work on $K$-PSI \cite{yehia_federated_K_PSI} achieves exponential communication complexity, which is significantly reduced in this work, as the download costs in Theorems  \ref{optimal downlaod cost k-psi-1}, \ref{optimal download cost fed-k-psi-1}, \ref{optimal download cost fed-k-psi-2} are all linear in the number of parties $M$, the number of databases per party $N$, and the number of colluding parties $T$.
\end{remark}

\section{Proposed Schemes}\label{proposed schemes}
The schemes proposed for both PMA and SPMA are based on CSA coding \cite{csa}, with further modifications to achieve symmetric privacy and security. In both problems, each party generates a private incidence vector $P_i$, $i \in [M]$. For the example shown in Fig.~\ref{sys_1} and Fig.~\ref{sys_2}, where $E=5$ and the alphabet is $\{a, b, c, d, e\}$, equivalently, $\{W_1,W_2,W_3,W_4,W_5\}$. Since $\mathcal{P}_1 = \{W_1,W_2,W_3,W_4,W_5\}$, the incidence vector of party 1 is $P_1=[1,1,1,1,1]^t$, since $\mathcal{P}_2 = \{W_2,W_3,W_4\}$, the incidence vector of party 2 is $P_2 = [0,1,1,1,0]^t$, and so on.    

\begin{remark}
Using the incidence vector to reply to the user's queries instead of the messages explicitly satisfies \eqref{parties_sec_1}.     
\end{remark}

\subsection{Proposed Scheme for PMA Type I}\label{scheme 1}
In PMA type I, there are $M$ parties, with $N$ databases each, out of which any $T$ can be colluding. Each party is allowed to eavesdrop on $Y$ communication links of any other party. Let the number of databases per party be $N= \max(T,Y)+1$. The vectors $P_i$, $i \in [M]$ are replicated in all the databases of each party. The user, who wishes to know how many times $W_{\theta}$ is repeated among the $M$ parties sends queries $Q_{ij}^{\theta}$,
\begin{align}\label{query_PSI}
    Q_{ij}^{\theta} = e_{\theta} +  \sum_{\ell=1}^{\mu}(1+\alpha_j)^{\ell} Z_{i\ell},
\end{align}
where $\mu = \max(T,Y)$, $e_{\theta}$ is a vector of length $E$ with 1 at the $\theta$th index and zeros otherwise, $Z_{i1}$s are independent noise vectors, with the same length, chosen uniformly at random, and $\alpha_j$s are globally known distinct constants. After receiving the queries, each database responds with an answer $A_{ij}^{\theta}$,
\begin{align}\label{Answer_PSI}
    A_{ij}^{\theta} = P_i^t Q_{ij}^{\theta} + S_{ij}, 
\end{align}
where $S_i =[S_{i1},\ldots, S_{iN}]^t$ is the masking vector corresponding to the $i$th party, unknown to the user. The masking vectors, $S_1, \ldots, S_M$, are chosen, independent of the incidence vectors, such that $\sum_{i=1}^M S_i = 0_N$, where $0_N$ is the zero vector of size $N\times1$. The answers from the $i$th party are given by,
\begin{align}
    A^{\theta}_i = \begin{bmatrix}
        A_{i1}, A_{i2}, \ldots,A_{iN}
    \end{bmatrix}^t = \Upsilon_N \Lambda_i + S_i,
\end{align}
where 
\begin{align}\label{xi_equation}
   \!\!\!\! \Upsilon_N= \begin{bmatrix}
        1&1+\alpha_1&(1+\alpha_1)^2& \ldots& (1+\alpha_1)^{N-1}\\
        1&1+\alpha_2&(1+\alpha_2)^2& \ldots& (1+\alpha_2)^{N-1}\\
        \vdots &\vdots&\vdots& &\vdots\\
        1&1+\alpha_N&(1+\alpha_N)^2& \ldots& (1+\alpha_N)^{N-1}\\
    \end{bmatrix},\!\!
\end{align}
and $ \Lambda_i = \begin{bmatrix}
        \mathbbm{1}(W_{\theta} \in \mathcal{P}_i),I_1, I_2,\ldots,  I_{N-1}
    \end{bmatrix}^t$ with $I_1$, $I_2, \ldots, I_{N-1}$ being interference symbols. To find the required answer, the user adds all the received answers, i.e.,
\begin{align}\label{decoding}
    \sum_{i=1}^M A_i^{\theta} =  \Upsilon_N \sum_{i=1}^M \Lambda_i=  \Upsilon  \begin{bmatrix}
        \sum_{i=1}^M \mathbbm{1}(W_{\theta} \in \mathcal{P}_i)\\
        \Tilde{I}_1\\
        \Tilde{I}_2\\
       \vdots\\
        \Tilde{I}_{N-1}
    \end{bmatrix},
\end{align}
and use the invertibility of $\Upsilon$ to obtain $\sum_{i=1}^M \mathbbm{1}(W_{\theta} \in \mathcal{P}_i)$.

\begin{remark}
    The number of databases per party required for the proposed scheme is $N\geq \max(Y,T)+1$ and the optimal number of databases per party that satisfies the minimum download cost, with a fixed $T$ and $Y$, for this scheme is $N=\max(T,Y)+1$.
\end{remark}

\begin{remark}
    In the proposed scheme given in this section, there is no exchange of information between parties except for the masking, which is independent of the messages and indices. Thus, \eqref{parties_sec_1}, \eqref{parties_sec_2} are both satisfied.
\end{remark}

\begin{remark}
    This scheme does not satisfy the symmetric privacy constraint in \eqref{fed_PSI_2} since the interference symbols may carry information about $\sum_{i}\mathbbm{1}(W_{\theta^C} \in \mathcal{P}_i)$. A modified version presented in Section~\ref{scheme 2} satisfies symmetric privacy.
\end{remark}

\begin{remark}
    The total communication complexity of the system, considering the sum of the user's upload cost, download cost, and the cost of sharing randomness between parties is given as $(M-1)N + EMN + MN$, where it is assumed that the masking vectors are generated by a single party, which are then sent to the rest of the parties.
\end{remark}

\subsection{Proposed Scheme for SPMA Type I }\label{scheme 2}
In this section, we assume that the number of databases in each party is $N=\max(T,Y)+1$, similar to the previous section, as the modification proposed for CSA to achieve symmetric privacy does not require additional databases. In contrast to PMA type I, SPMA type I hides any information about the availability of messages other than the one being checked from the user. Intuitively, if the parties can utilize random noise in the scheme such that the noise hides the contents of the interference symbols in \eqref{decoding}, the scheme becomes private in both directions. The core difference between this scheme and the scheme in Section~\ref{scheme 1} is that the databases within each party in this scheme share common randomness $Z'$ which is generated independently from the messages, the incidence vector, and the masking variables. As in the previous section, the $i$th party stores its incidence vector $P_i$ in a replicated manner in all $N$ databases, i.e.,
\begin{align}
    P_{ij} = P_{i}, \quad i \in [M], ~j \in [N]. 
\end{align}
The user sends the same queries as in \eqref{query_PSI}, to which the databases send the corresponding answers given by,
\begin{align}
    A_{ij}^{\theta} = P_{ij}^t Q_{ij}^{\theta} + \sum_{\ell=1}^ {N-1} (1+\alpha_j)^{\ell} Z'_{i\ell} +S_{ij},
\end{align}
where $Z'_{i\ell}$s are random noise variables initialized and shared by the $N$ databases in each party $i$. Thus, the user obtains,
\begin{align}
    A^{\theta}_i= \Upsilon_N \Lambda_i +S_i,
\end{align}
where $\Upsilon$ is the same as in \eqref{xi_equation}, $S_i$ is the masking vector, and $\Lambda_i$ is given by,
\begin{align}
    \Lambda_i = \begin{bmatrix}
        \mathbbm{1}(W_{\theta} \in \mathcal{P}_i)\\
        I_1 + Z'_{i1}\\
        I_2+ Z'_{i2}\\
       \vdots\\
        I_{N-1}+Z'_{i,N-1}
    \end{bmatrix},
\end{align}
where $I_1$, $I_2, \ldots I_{N-1}$ are interference symbols. By applying the same decoding scheme as in the previous section, the user retrieves the required answer.

\begin{remark}
    The total communication complexity of this scheme is given by $(M-1)N + EMN+ N-1 + MN  $, where we assume that one party generates the masking vectors and sends them to the rest of the parties.
\end{remark}

\begin{remark}
    If, in addition, we assume that $T_2$ parties are communicating, i.e., sharing their datasets to figure out the datasets of the remaining $M-T_2$ parties, the schemes presented in the previous sections still maintain the same download cost since the schemes do not require the parties to share their datasets, nor the incidence vectors.
\end{remark}

\subsection{Proposed Scheme for SPMA Type II}\label{scheme 3}
In this case, there are $M$ parties, each with $N$ databases, and all databases in $T < M$ parties are allowed to collude. The main issue here is that the efficient PIR schemes cannot be applied separately to each party, as all the databases in each party collude with each other. This requires any information exchange among the parties to be secure against any $N$ communicating databases, which motivates the mapping of this problem to an XSTPIR problem \cite{csa} with the number of colluding databases, i.e., databases that share information of the users, $T'=NT$, and the number of communicating databases, i.e., databases that share their contents, $X=N$. We adopt a variant of the CSA scheme to solve this problem as in the previous sections. For this scheme, $N$ is chosen such that $MN = N+\max(TN,Y_1,\ldots, Y_M) + 1$, and the proposed approach is defined in the following steps: 

\subsubsection*{Step 1: Initialization and Distribution}
Each party with its message set, $\mathcal{P}_i$, has its corresponding incidence vector $P_i$ that needs to be secure against any $N$ communicating databases. Thus, it is encoded as,
\begin{align}\label{secure_incidence}
    \Tilde{P}_{ij} = P_i + \sum_{\ell=1}^N(1+\alpha_j)^{\ell}X_{i\ell}, ~i \in [M],~ j \in [MN],
\end{align}
where $X_{i\ell}$s are independent random noise vectors. The $i$th party sends $\Tilde{P}_{ij}$ to the $j$th database. After receiving $\Tilde{P}_{ij}, ~i \in [M]$, the $j$th database adds all the received vectors as, 
\begin{align}
    \Tilde{P}_j  = \sum_{i=1}^M P_i + (1+\alpha_j)\Tilde{X}_1 + \ldots + (1+\alpha_j)^N \Tilde{X}_N, 
\end{align}
for $j \in [MN]$, where $\Tilde{X}_n = \sum_{i=1}^M X_{in}, ~n \in [N]$. 

\subsubsection*{Step 2: Queries and Answers Structure}
The user who wants to know how many times $W_{\theta}$ is repeated in the $M$ parties, sends the following query to the $n$th database, 
\begin{align}
    Q_{n}^{\theta} = e_{\theta} + \sum_{\ell=1}^{\mu}  (1+\alpha_n)^{\ell}Z_{\ell},~ n \in [MN],
\end{align}
where $Z_k,~ k\in [\mu]$ are uniform independent noise vectors, and $\mu = \max(NT,Y_1,\ldots,Y_M)$. The parties agree on uniform random noise variables $Z'_1,Z'_2,\ldots,Z'_{MN-1}$, and generate the answers to achieve symmetric privacy as,
\begin{align}
    A_n^{\theta} = \Tilde{P}_n^t Q_n^{\theta} + \sum_{i=1}^{MN-1} (1+\alpha_n)^i Z'_i,\quad n\in[MN].
\end{align}
\subsubsection*{Decoding Structure}
After retrieving all the answers, the user has the following answer vector
\begin{align}
A^{\theta} = \begin{bmatrix}
    A_1^{\theta}\\
    A_2^{\theta}\\
   \vdots\\
    A_{MN}^{\theta}
\end{bmatrix}=\Upsilon_{MN} \begin{bmatrix}
    \sum_{i=1}^M \mathbbm{1}(W_{\theta} \in \mathcal{P}_i)\\
    I_1 + Z'_1\\
    \vdots\\
    I_{MN-1}+ Z'_{MN-1}
\end{bmatrix}, 
\end{align}
where $I_1,\ldots,I_{MN-1}$ are the interference symbols. The user multiplies the answer vector by $\Upsilon_{MN}^{-1}$ to obtain the required information. The download cost in the proposed scheme is $MN$ with $MN \geq N+TN+1$, which concludes the proof of the upper bound in Theorem~\ref{optimal download cost fed-k-psi-2}.

\begin{remark}
   The total communication cost in this scheme is equal to $(E+1)(N+TN+1) + N+NT$.
\end{remark}

\begin{remark}
    If $MN > N+\max(TN,Y_1,\ldots,Y_M)+1$, we can drop the extra databases. Interestingly, this shows that cooperation between parties, even though their datasets are secure from each other, can save some databases.
\end{remark}

\begin{remark}
    In this scheme, if there are $T_2$ communicating parties, then the optimal download cost is upper bounded by $T_2N+ \max(TN,Y_1,\ldots,Y_M) + 1$. 
\end{remark}

\section{Conclusions}\label{conclusions}
In this paper, we introduced PMA which is a generalization and refinement of $K$-PSI. In PMA, the user wishes to know how many times a certain message appears in all parties. We consider different cases of the problem, based on the privacy requirements (user-privacy and symmetric privacy) and database collusions. We proposed achievable schemes for all cases considered, focusing on the behavior of the communication complexity as a function of system parameters. Compared to the previous work in $K$-PSI that achieves \emph{exponential} complexity, the schemes proposed here achieve \emph{linear} complexity with enhanced privacy and security guarantees.

\section{Appendix: Proofs}\label{proofs}
In this section, we prove important lemmas that collectively prove the security and privacy requirements for the developed schemes. More precisely, Lemmas~\ref{masking_lemma}, \ref{colluding lemma}, and \ref{eavs_1_lemma} prove Theorem~\ref{optimal downlaod cost k-psi-1}; Lemmas \ref{symmetry_lemma}, \ref{masking_lemma}, \ref{colluding lemma}, and \ref{eavs_1_lemma} prove Theorem~\ref{optimal download cost fed-k-psi-1}; and Lemmas~\ref{symmetry_lemma}, \ref{security_lemma}, \ref{colluding lemma}, and \ref{eavs_2_lemma} prove Theorem~\ref{optimal download cost fed-k-psi-2}, in terms of the claims made on the levels of privacy and security achieved.  

\begin{lemma}\label{symmetry_lemma}
    The schemes proposed for SPMA type I and SPMA type II provide symmetric privacy.
\end{lemma}

\begin{Proof}
    For the scheme proposed for the SPMA type I problem, since every party has its own independent random variables, $Z'_1(i), \ldots,Z'_{N-1}(i), ~i \in [M]$, it suffices to consider each party independently. Let $\Gamma_{\theta^C}(i) = \bigcup_{k=1\atop{k\neq \theta}}^E\{\mathbbm{1}(W_k \in \mathcal{P}_i)\}$, thus we need to show that $I(\Gamma_{\theta^C}(i);A_{i[N]}^{\theta}(i)|Q_{i[N]}^{\theta}(i),\kappa^{\theta})=0$. The answers received by the user from the $i$th party are contaminated with noise terms independent of the messages along the interference terms, which is simply a random noise symbol $Z'$ unknown to the user. Thus,
    \begin{align}
    &I(\Gamma_{\theta^C}(i); A^{\theta}_{i[N]}|\kappa^{\theta},Q_{[MN]}^{\theta},\theta)\nonumber\\
    &=H(\Gamma_{\theta^C}(i)|\kappa^{\theta},Q_{[MN]}^{\theta},\theta)\nonumber\\
    &\quad -H(\Gamma_{\theta^C}(i)|A^{[\theta]}_{i[N]},\kappa^{\theta},Q_{[MN]}^{\theta},\theta)\\
    &\leq H(\Gamma_{\theta^C}(i)) - H(\Gamma_{\theta^C}(i)|\kappa^{\theta},Z'_{[N-1]},Q_{[MN]}^{\theta},\theta)\\
    &= 0.
    \end{align}
For the scheme presented for the SPMA type II, the same approach is used, however, the interference terms in the answers are contaminated with independent random noise symbols collectively, thus we use $Z'_1, \ldots,Z'_{MN-1}$ in the proof and consider the answers collectively.  
\end{Proof}

\begin{lemma}\label{masking_lemma}
    The masking used in PMA type I and SPMA type I schemes guarantees blind estimation requirements \eqref{PSI_2}-\eqref{PSI_3}.
\end{lemma}
\begin{Proof}
Assume that the random vector $\Omega(W_{\theta})$ represents the information that the user requires about $W_{\theta}$ and any possible side information about the presence of the same message in a subset of parties $\mathcal{M}$ such that $|\mathcal{M}| \leq M-1$. Let $A^{\theta}_{\mathcal{M}}$ be the set of answers received from those parties and $Z$ be the noise terms used to ensure user privacy, then
\begin{align}
   I(\Omega(W_{\theta}); A_{\mathcal{M}}^{\theta}|\theta, Z)&\!=\!I(\Omega(W_{\theta}); A_{i_1}^{\theta},\ldots,A^{\theta}_{i_{M-1}})\\
   &\!=\!I(\Omega(W_{\theta});S_{i_1},\ldots,S_{i_{M-1}})= 0
\end{align}
where the last equality is due to the independence between the incidence vectors and the masking vectors.

Now, since the answer vectors from each party, $A_1,\ldots, A_M$, are aligned, the answers of all $M$ parties cannot give any information about any subset of parties with cardinality less than $M$.
\end{Proof}

\begin{lemma}\label{security_lemma} 
    The scheme proposed for SPMA type II is secure against any $N$ communicating databases.
\end{lemma}

\begin{Proof}
    Let $P \vcentcolon = \Tilde{P}_j$ given in \eqref{secure_incidence} for any $j$, and $P(\ell)$ is the $\ell$th element of $P$. Define the vector of any $N$ observations for the $\ell$th element of $P$ as $U_\ell$. Then, $U_\ell$ can be written as
\begin{align}
U_\ell&=[U_\ell(1),\ldots,U_\ell(N)]^t \\
&=P(\ell)\begin{bmatrix}
    1\\ \vdots\\ 1
\end{bmatrix} + \begin{bmatrix}
    1+\alpha_{i_1}&\ldots&(1+\alpha_{i_1})^N\\
    1+\alpha_{i_2}&\ldots&(1+\alpha_{i_2})^N\\
    \vdots& &\vdots\\
    1+\alpha_{i_N}&\ldots&(1+\alpha_{i_N})^N
\end{bmatrix}\begin{bmatrix}
    X_1(\ell)\\
    \vdots\\
    X_N(\ell)
\end{bmatrix}\\
&= P(\ell)\mathbf{1} +\diag(\mathbf{1}+\alpha_{[i_1:i_N]}) ~\Upsilon_N[ X_1(\ell),\dotsc,X_N(\ell)]^t\label{eq1}.
\end{align}
 In \eqref{eq1}, the matrices $\diag(\mathbf{1}+\alpha_{[i_1:i_N]})$ and $\Upsilon_N$ are invertible, which makes their product invertible as well. Then, the following inequalities hold
 \begin{align}
 I&(P(\ell);U_\ell)\nonumber\\&=I(P(\ell);P(\ell)\mathbf{1}+ \diag(1+\alpha_{[i_1:i_N]}) ~\Upsilon_N X(\ell))\\
 &= I(P(\ell);P(\ell)(\diag(1+\alpha_{[i_1:i_N]}) ~\Upsilon_N)^{-1}\mathbf{1}+X(\ell))\\ 
 &= I(P(\ell);X(\ell))=0,
 \end{align}
where $X(\ell) =[X_1(\ell),\ldots,X_N(\ell)]^t$. 
\end{Proof}

\begin{lemma}\label{colluding lemma}
    The query structure defined in the schemes for PMA type I, SPMA type I and SPMA type II are secure against any $T$ colluding databases.
\end{lemma}

\begin{Proof}
    Let $\theta$ be the required user index, then for any set of colluding servers $\mathcal{T}$ such that $ |\mathcal{T}|\leq T$, the collective observations $Q_{\mathcal{T}}$ can be written as
    \begin{align}
        Q_{\mathcal{T}} &= [Q_{i_1}^t,\ldots,Q_{i_T}^t]= E_{\theta} + B\begin{bmatrix}
    Z_1^t,\ldots,Z_T^t    
\end{bmatrix}^t,
    \end{align}
    where $E_{\theta} =[e_{\theta}^t,\ldots,e_{\theta}^t]^t$. Note that $H(\theta|E_{\theta})=0$, and the matrix $B=\diag(\mathbf{1}+\alpha_{[i_1:i_T]}) \Upsilon_T$ is invertible as proven in Lemma~\ref{security_lemma}. Now, to ensure privacy given the queries $Q_{\mathcal{T}}$, we proceed as,
\begin{align}
    I(\theta;Q_{\mathcal{T}})&= I(\theta;E_{\theta}+BZ) \\
    &= I(\theta;B^{-1}E_{\theta}+Z) = I(\theta;Z)=0,
\end{align} 
which concludes the proof.
\end{Proof}

\begin{lemma}\label{eavs_1_lemma}
    The schemes proposed for PMA type I and SPMA type I are secure against an eavesdropper who has access to any $\mathcal{Y}$, where $|\mathcal{Y}| \leq Y$ answers from the other parties.
\end{lemma}

\begin{Proof}
    We note that the incidence vectors for each party are independent from each others, i.e., $I(P_i;P_\ell) = 0, ~ i \neq \ell$. Thus, $I(A_{i\mathcal{Y}};P_i|P_\ell)=I(A_{i\mathcal{Y}};P_i)$. Now, using the same proof as in Lemma~\ref{colluding lemma} with $\mathcal{Y}$ instead of $\mathcal{T}$ and $P_i$ instead of $\theta$, the proof follows.
\end{Proof}

\begin{lemma}\label{eavs_2_lemma}
    The scheme proposed for SPMA type II protects the contents of the answers from any party that eavesdrops on $\mathcal{Y}_i$ links of other parties. 
\end{lemma}

\begin{Proof}
    First, note that $I(S_n; \mathcal{P}_i) = 0$, where $S_n$ is the storage in the $j$th database. This means that the dataset is secure against any party. Let $\mathcal{Y}= \max(\mathcal{Y}_i)$. Thus,
    \begin{align}
        I(\sum_{i=1}^M P_i;A_{\mathcal{Y}}|Q_{\mathcal{Y}}) &\!=\!I(\sum_{i=1}^M P_i;A_{\mathcal{Y}}) \\ &\!=\!I(e_{\theta}^t\sum_{i=1}^M P_i;A_{\mathcal{Y}})+I(\Gamma^{\theta};A_{\mathcal{Y}})\label{chain_rule_eavesdropper_lemma}=0
    \end{align}
    where $A_{\mathcal{Y}}$ are the answers from any $\mathcal{Y}$ databases. The first term on the right hand side of \eqref{chain_rule_eavesdropper_lemma} is equal to zero by the same method as in the proof of Lemma~\ref{colluding lemma}, where we replace the queries with the answers and $\theta$ with $e_{\theta}^t\sum_{i=1}^M P_i$. The second term is equal to zero as a direct consequence of Lemma~\ref{symmetry_lemma}.
\end{Proof}

\bibliographystyle{unsrt}
\bibliography{references}
\end{document}